\documentstyle[12pt,epsf,here]{article}
\def \be {\begin{equation}}
\def \e {\end{equation}}
\def \bea {\begin{eqnarray}}
\def \ea {\end{eqnarray}}

\def \no {\nonumber}

\def \sub {\scriptscriptstyle}
\newcommand{\To}[2]{\stackrel{#1}{\hbox to #2 pt{\rightarrowfill}}}

\def\np#1#2#3{{\it Nucl.~Phys.\/}~{\bf B#1} (19#2) #3}
\def\pl#1#2#3{{\it Phys.~Lett.\/}~{\bf B#1} (19#2) #3}

\def\prd#1#2#3{{\it Phys.~Rev.\/}~{\bf D#1} (19#2) #3}

\def\zpc#1#2#3{{\it Z.~Phys.\/}~{\bf C#1} (19#2) #3}

\begin{document}

\title{\vspace*{-3cm}
\begin{minipage}[l]{12cm}
$\quad$
\end{minipage}
\begin{minipage}[r]{3cm}
{\small PITHA 98/38}\\
\noindent
{\small April 1999}\\ {\small Revised} \\  $\quad $\\
\end{minipage}
\noindent
\bf Angular Distribution of Decay Leptons from \boldmath{$e^+e^- \to W^+W^-$} at Threshold}
\author{A. Werthenbach\thanks{email: Anja.Werthenbach@durham.ac.uk} \\ {Department of Physics, University of Durham, UK} \and   L.~M. Sehgal\thanks{email: sehgal@physik.rwth-aachen.de} \\ { Institute of Theoretical Physics (E), RWTH Aachen, Germany}}
\date{}
\maketitle

\vspace{1cm}

\abstract{The reaction $ e^+e^- \to W^+W^-$ produces a $W$-boson pair with a non-trivial spin correlation at threshold. This correlation leads to a characteristic angular correlation between the leptons produced in $W^{\pm} \to \ell ^{\pm} \nu$ (angles relative to the $e^-$ beam direction):  $d \sigma /d \, \cos \theta_+ d \, \cos \theta_- \sim (1-\cos \theta_+) ( 1+ \cos \theta_-) (1+\cos \theta_- \cos \theta_+)  \,$ . If only the $\ell^-$ is observed, its angular distribution is $d \sigma / d \cos \theta_- \sim (1+\cos \theta_-)(3-\cos \theta_-) \,$, implying a forward-backward asymmetry of $3/8$. An analytic result is also given for the azimuthal correlation. These results are reproduced by a Monte Carlo program, that also enables us to study the effects of the $W$ decay width. The threshold behaviour, which stems from the dominance of $\nu$-exchange, is contrasted with that due to $\gamma$- and $Z$-exchange, which is relevant for annihilation in the helicity state $e^-_Re^+_L$.\\}

\newpage

\noindent
We consider in this paper some simple but non-trivial aspects of the reaction $e^+e^- \to W^+W^-$ at threshold, which manifest themselves in the angular distribution of the decay leptons produced by $W^+ \to \ell^+ \nu_{\ell}, W^- \to \ell^- \bar{\nu}_{\ell}$. These features may be of interest in measurements at $\sqrt{s} \simeq 2M_W $ which have as their primary aim the determination of the $W$-mass \cite{massw,james1,experiment}. We give analytic results valid at threshold which can be compared with the characteristics of leptons in final states of the type $\ell^+\ell^-\nu_{\ell} \bar{\nu}_{\ell} $, $\, \ell^- \bar{\nu}_{\ell} q \bar{q}^{\prime} $ or $ \ell^+ {\nu}_{\ell} q \bar{q}^{\prime} $.\\

\noindent
At threshold $( \beta = \left(1-{4M_W^2}/{s}\right)^{1/2} \to 0 )$ the reaction $e^+e^- \!\to \!W^+W^-$ is dominated by annihilation in the helicity state $e^-_L e^+_R$, with the cross section
\bea
\label{L}
\left( \frac{ d \sigma}{d \cos \theta}  \right)_{ e^-_L e^+_R} = \beta \, \frac{e^4}{8 \pi \sin ^4 \theta_w} \, \frac{1}{s} \quad ,
\ea
\noindent
$\theta$ being the production angle of the $W^-$ relative to the $e^-$ beam direction. This isotropic cross section is determined by the $\nu$-exchange diagram, corrections due to $\gamma$- and $Z$-exchange being ${\cal O}(\beta^2)$. 
The isotropy implies that the $W^+W^-$ system is produced in an $S$-wave state $L=0$. This, together with the fact that the annihilating state is $e_L^-e_R^+$, means that the quantum numbers of the $W^+W^-$ system are $S=1, S_z=-1$, where the $z$-axis is along the $e^-$ beam direction. The spin wave function of the $W^+W^-$ pair is therefore

\bea
|W^-W^+; S=1,S_z=-1 \rangle &=& \frac{1}{\sqrt{2}} [ |W^-(S_z=-1)\rangle\,\, |W^+(S_z=0)\rangle \no \\
&& \hspace{2cm}-  |W^-(S_z=0)\rangle \,\,|W^+(S_z=-1)\rangle ] \; .
\ea
Now, the decay amplitudes of the $W^-$ and $W^+$ are (up to a normalization factor)
\bea
 W^-(S_z=-1) \to \ell^- \bar{\nu}_{\ell} \, &:& \, (1+ \cos \theta_-)\, e^{-i\phi_-} \no \\
 W^-(S_z=0) \to \ell^- \bar{\nu}_{\ell} \, &:& \,  \sin \theta_-  \no \\
 W^+(S_z=-1) \to \ell^+ {\nu}_{\ell} \,& :& \, (1- \cos \theta_+)\, e^{-i\phi_+} \no \\
 W^+(S_z=0) \to \ell^+ {\nu}_{\ell} \, &:& \,  \sin \theta_+  
\ea
where $(\theta_+,\phi_+)$ and $(\theta_-, \phi_-)$ are the polar and azimuthal angles of the secondary leptons $\ell^+$ and $\ell^-$, respectively. Hence the amplitude for $ |W^-W^+ \rangle \to |\ell^- \bar{\nu}_\ell; \ell^+ \nu_{\ell}\rangle$ is
\bea
{\cal A} = const. \,\, [  (1+ \cos \theta_-)\, e^{-i\phi_-}  \sin \theta_+ -  (1- \cos \theta_+)\, e^{-i\phi_+} \sin \theta_-  ] \; .
\ea
The corresponding angular distribution is
\bea
\label{angular}
\frac{d \sigma}{d \cos \theta_+ d \cos \theta_- d \phi_+ d \phi_-}&=& const.\,\, \big[ (1+ \cos \theta_-)^2  \sin^2 \theta_+ + (1- \cos \theta_+)^2  \sin^2 \theta_- \no \\
&& \hspace{-0.9cm} - 2  (1+ \cos \theta_-) (1- \cos \theta_+) \sin \theta_- \sin \theta_+\cos(\phi_--\phi_+) \big] \; .
\ea
Integrating over the azimuthal angles, we obtain the polar angle correlation of the two leptons:
\bea
\label{l2dim}
\left( \frac{d \sigma}{d \cos \theta_+ d \cos \theta_- } \right)_{ e^-_L e^+_R} \!\! = const. \,\,  (1- \cos \theta_+)  (1+ \cos \theta_-)(1+ \cos \theta_- \cos \theta_+)
\ea

\noindent
which is plotted in Fig.~1. The zero at $\cos \theta_-=-1$ implies that an $\ell^- $ cannot be produced exactly backwards; likewise the zero at $\cos \theta_+ =+1$ forbids an $\ell^+$ in the forward direction. Both features are consequences of the $V-A$ structure of the $\nu$-exchange diagram and the $S$-wave character of the $W^+W^-$ state. If only one decay lepton is observed (for example in semileptonic final states $\ell^- \bar{\nu}_{\ell} q \bar{q}^{\prime}$ ), the angular distribution is

\bea
\label{l1dim}
\left( \frac{d \sigma}{d \cos \theta_-} \right)_{ e^-_L e^+_R} = const. \,\,(1+\cos \theta_-)(3-\cos \theta_-)  \quad .
\ea

\noindent
This implies, in particular, that an $\ell^-$ produced at threshold should have a forward-backward asymmetry of 3/8. The result (\ref{l1dim}) was also obtained as a limiting case in \cite{rekalo,james2}.\\

\noindent
It is possible also to obtain from Eq.~(\ref{angular}) the correlation in the relative azimuth of the two leptons. Defining $\Delta=\frac{1}{2} (\phi_--\phi_+), \Sigma = \frac{1}{2} (\phi_-+\phi_+)$, and noting that the range of the variables is
\bea
0\, & \leq &\, |\Delta| \, \leq \, \pi \no \\
| \Delta |\, &\leq & \, \Sigma \leq \,2 \pi - | \Delta | \; ,
\ea
integration over $\cos \theta_+, \cos \theta_-$ and $ \Sigma$ gives the result
\bea
\label{azi}
\frac{d \sigma}{d | \Delta|} = const. \,\, (1- \frac{|\Delta|}{\pi}) ( 1- \frac{9 \pi^2}{128} \cos 2 |\Delta|),
\ea
which is plotted in Fig.~2.\\

\noindent
It is remarkable that the isotropic threshold cross section Eq.~(\ref{L}) actually results from a superposition of various terms $ d \sigma ^{\lambda_- \lambda_+}/ d \cos \theta$ describing production of the helicity states $W^-(\lambda_-) W^+(\lambda _+)$, which individually are not isotropic, even in the limit $\beta \to 0$. These helicity cross sections are listed in the Appendix. By combining the helicity amplitudes for $ e^-e^+ \to W^-W^+$ at threshold with the $W^{\pm}$ decay amplitudes \cite{gounaris,hagi}, and integrating over the $W$ production angle, we have verified the polar and azimuthal correlation given in Eqs.(\ref{l2dim}) and (\ref{azi}) above.\\

\noindent
In order to see how the threshold result Eq.~(\ref{l1dim}) is affected by the non-zero decay width of the $W$, we have used a Monte Carlo program \cite{james2} describing the 
reaction $e^+e^- \to W^+W^-$ followed by the decay $W^- \to \ell^-\bar{\nu_{\ell}}$. The effects of the decay width were simulated by assuming the mass of the $W$ to be distributed according to a Breit-Wigner function. The results are shown in Fig.~3. In the zero-width limit, the Monte Carlo program reproduces the analytic result given in Eq.~(\ref{l1dim}). On the other hand, with $\Gamma_W \approx 2 GeV$, the angular distribution is distorted towards a larger forward-backward asymmetry, $A_{FB}=0.45$. It is possible that the variation of the forward-backward asymmetry in the threshold region would be a useful adjunct to the threshold behaviour of the total cross section $\sigma(e^+e^- \to W^+W^-)$ as a probe of $M_W$ and $\Gamma_W$. \\

\noindent
Since the threshold results given in Eq.~(\ref{l2dim}),~(\ref{l1dim}) and (\ref{azi}) stem entirely from the neutrino exchange diagram, there is some theoretical interest in the question what type of threshold behaviour would result from $\gamma$- and $Z$-exchange. One way to isolate these diagrams is to consider the process $e^+e^- \to W^+W^-$ with $e^+e^-$ beams polarized in the state $e^-_Re^+_L$. The threshold behaviour of $e^-_Re^+_L \to W^-W^+$ is found to be 
\bea
\label{R}
\left( \frac{d \sigma}{d \cos \theta}\right)_{e^-_Re^+_L} = \frac{e^4}{32 \pi }s \beta^3 \left( \frac{1}{s}-\frac{1}{s-M_Z^2} \right)^2 (19 - 3\cos ^2 \theta)
\ea

\noindent
This cross section is not isotropic, and the threshold behaviour $\beta^3$ is indicative of an $L=1$ state
. Making use of the helicity amplitudes for $e^-_Re^+_L \to W^-(\lambda_-)W^+(\lambda_+)$, the correlation between the polar angles of the secondary leptons turns out to be (using the abbreviation $\cos \theta_- \equiv y, \cos \theta_+ \equiv z):$

\bea
\label{r2dim}
\left( \frac{d \sigma}{d y dz} \right)_{e^-_Re^+_L} &=& const.\, \left(24z-8yz^2+19-28yz \right.  \no \\
&&  \hspace{1cm}  \left. +8y^2z-24y+7z^2+7y^2+3y^2z^2\right) \, .
\ea

\noindent
which is plotted in Fig.~4.  The angular distribution of a single lepton $\ell ^-$ is

\bea
\label{r1dim}
\left( \frac{d \sigma}{d \cos \theta_-} \right)_{e^-_Re^+_L} =  const. \,(4-3\cos \theta_-)(2-\cos \theta_-)
\ea

\noindent
implying that the $\ell^-$ is produced predominantly in the backward hemisphere, the forward-backward asymmetry being $-5/9$.\\
 
\noindent
The azimuthal correlation, analogous to Eq.~(\ref{azi}), is found to be
\bea
\label{razi}
\frac{d \sigma}{d |\Delta|} &=& const.\,\, \Big[ \sin 4 |\Delta| - \frac{9 \pi ^2}{64}\sin 2 | \Delta | \no \\
&& \hspace{1cm}+ \frac{\pi}{4} (1-\frac{|\Delta|}{\pi}) \left( 54 + \cos 4 |\Delta| -\frac{27 \pi^2}{16} \cos 2 |\Delta| \right) \Big]  
\ea
\noindent
and illustrated in Fig.~5.\\

\noindent
We have also used the Monte Carlo program to check the result (\ref{r1dim}) describing $e_R^-e_L^+$ annihilation. As shown in Fig.~6, the program reproduces the angular distribution (\ref{r1dim}), and provides an estimate of the distortion due to non-zero $\Gamma_W$.\\

\noindent
In principle, the results given in Eqs.~(\ref{l2dim}) and (\ref{azi}) can be compared with data obtained in the LEP experiments at c.m.-energy $\sqrt{s}=161$~GeV. However, the statistics available in the $\ell^+\nu  \ell^- \bar{\nu}$ channel, required for testing this correlation, are limited to about a dozen events. On the other hand, the forward-backward asymmetry contained in the single lepton distribution Eq.~(\ref{l1dim}) can be tested using the final states $\ell^-\bar{\nu}_{\ell} q \bar{q} $ and $\ell^+\nu_{\ell} q \bar{q} $, for which an order of magnitude more events are available. There is every reason to believe that the reaction  $e^-e^+ \to W^-W^+$ will continue to receive close scrutiny at future $e^+e^-$ colliders, and we believe that the threshold results given in this paper will provide a useful guideline.

\newpage
\vspace*{-0.7cm}
\noindent
{\bf Acknowledgements}\\

\noindent
One of us (L.~M.~S) would like to acknowledge useful discussions with Prof. S.~Rindani and Dr. P. Poulose during a visit to the Physical Research Laboratory, Ahmedabad.\\
\noindent
A.~W gratefully acknowledges financial support in the form of a 'DAAD Doktorandenstipendium im Rahmen des gemeinsamen Hochschulprogramms III f\"ur Bund und L\"ander'. AW is also grateful to the Institute of Theoretical Physics (E), RWTH Aachen, for hospitality during the completion of this work. \\

\vspace{0.3cm}
\noindent
{\bf Appendix}\\

\noindent
Helicity cross sections at threshold for $e^-_L e^+_R \to W^-(\lambda_-)W^+(\lambda_+)$
\bea
\frac{d \sigma^{00}}{d \cos \theta } &= & \frac{1}{8 \, \pi \, s}\, \frac{e^4 \, \beta \, (\sin ^2 \theta \, \cos ^2 \theta) }{\sin ^4 \theta _{\sub{W}}} \no \\
\frac{d \sigma^{++}}{d \cos \theta } &= &\frac{d \sigma^{--}}{d \cos \theta } = \frac{1}{32 \, \pi \, s}\, \frac{e^4 \, \beta \, (\sin ^2 \theta \, \cos ^2 \theta) }{\sin ^4 \theta _{\sub{W}}} \no \\
\frac{d \sigma^{+0}}{d \cos \theta } &=&  \frac{d \sigma^{0-}}{d \cos \theta } =  \frac{1}{64 \, \pi \, s}\, \frac{e^4 \, \beta \,[(-2\,\cos^2 \theta +1)-\,\cos \theta ]^2  }{\sin ^4 \theta _{\sub{W}}} \no \\
\frac{d \sigma^{-0}}{d \cos \theta } &=&  \frac{d \sigma^{0+}}{d \cos \theta } =  \frac{1}{64 \, \pi \, s}\, \frac{e^4 \, \beta \,[(-2\,\cos^2 \theta +1)+\,\cos \theta ]^2  }{\sin ^4 \theta _{\sub{W}}} \no \\
\frac{d \sigma^{+-}}{d \cos \theta } &=&  \frac{1}{32 \, \pi \, s}\, \frac{e^4 \, \beta \sin ^2 \theta\, (\cos \theta +1)^2 }{\sin ^4 \theta _{\sub{W}}} \no \\
\frac{d \sigma^{-+}}{d \cos \theta } &=&  \frac{1}{32 \, \pi \, s}\, \frac{e^4 \, \beta \sin ^2 \theta\, (\cos \theta -1)^2 }{\sin ^4 \theta _{\sub{W}}} \:, \no \\
\ea
\noindent
and for  $e^-_Re^+_L \to W^-(\lambda_-)W^+(\lambda_+)$
\bea
\frac{d\sigma ^{00}}{d \cos \theta} & = & \frac{9}{32\pi s} \left( \frac{e^2M^2_Z}{s-M^2_Z} \right) ^2 \beta^3 \sin ^2 \theta \no\\
\frac{d\sigma ^{++}}{d \cos \theta} & = & \frac{d\sigma ^{--}}{d \cos \theta} = \frac{1}{32\pi s} \left( \frac{e^2M^2_Z}{s-M^2_Z} \right) ^2\beta^3  \sin ^2 \theta \no \\
\frac{d\sigma ^{+0}}{d \cos \theta} & = & \frac{d\sigma ^{0-}}{d \cos \theta}= \frac{1}{16\pi s}  \left( \frac{e^2M^2_Z}{s-M^2_Z} \right) ^2 \beta^3 (1-\cos \theta )^2 \no \\
\frac{d\sigma ^{-0}}{d \cos \theta} & = &\frac{d\sigma ^{0+}}{d \cos \theta}= \frac{1}{16\pi s} \left( \frac{e^2M^2_Z}{s-M^2_Z} \right) ^2 \beta^3  (1+\cos \theta)^2 \no \\
\frac{d\sigma ^{+-}}{d \cos \theta} & = &\frac{d\sigma ^{-+}}{d \cos \theta} = 0 \: .
\ea

\newpage

\noindent
{\bf Figure captions}
\vspace{0.5cm}

\noindent
\begin{minipage}[l]{2cm}Figure 1: \\ \quad \end{minipage}
\begin{minipage}[r]{12cm}{Polar angle correlation $d \sigma / d \cos \theta_+ d \cos \theta_-$ of leptons in $e^-_L e^+_R \to W^-W^+$ at threshold.}\end{minipage}\\

\vspace{0.3cm}
\noindent
\begin{minipage}[l]{2cm}Figure 2: \\ \quad \end{minipage}
\begin{minipage}[r]{12cm}{{Azimuthal correlation $d \sigma / d |\Delta|$ of leptons in $e^-_L e^+_R \to W^-W^+$ at threshold ($ \Delta= \frac{1}{2} (\phi_- - \phi_+)$.}}\end{minipage}\\

\vspace{0.3cm}
\noindent
\begin{minipage}[l]{2cm}Figure 3: \\ \quad \\ \quad \end{minipage}
\begin{minipage}[r]{12cm}{{Single lepton distribution $d \sigma / d \cos \theta_-$ from  $e^-_L e^+_R \to W^-W^+$ obtained by Monte Carlo for $\Gamma_W \approx 0$ and $\Gamma_W=2.06$~GeV, compared with analytic result Eq.~(\ref{l1dim}).}}\end{minipage}  \\

\vspace{0.3cm}
\noindent
\begin{minipage}[l]{2cm}Figure 4:  \end{minipage}
\begin{minipage}[r]{12cm}{{ Same as Fig.~1 for $e^-_R e^+_L \to W^-W^+$.}}\end{minipage}  \\

\vspace{0.3cm}
\noindent
\begin{minipage}[l]{2cm}Figure 5:  \end{minipage}
\begin{minipage}[r]{12cm}{ { Same as Fig.~2 for $e^-_R e^+_L \to W^-W^+$.}}\end{minipage}\\

\vspace{0.3cm}
\noindent
\begin{minipage}[l]{2cm}Figure 6:  \\ \quad   \end{minipage}
\begin{minipage}[r]{12cm}{ { Same as Fig.~3 for $e^-_R e^+_L \to W^-W^+$ compared with analytic result Eq.~(\ref{r1dim}) .}}\end{minipage}\\

\newpage

\vspace*{-3.5cm}
\begin{minipage}[l]{5cm}
\begin{figure}[H]
\centerline{\epsfysize=10cm\epsffile{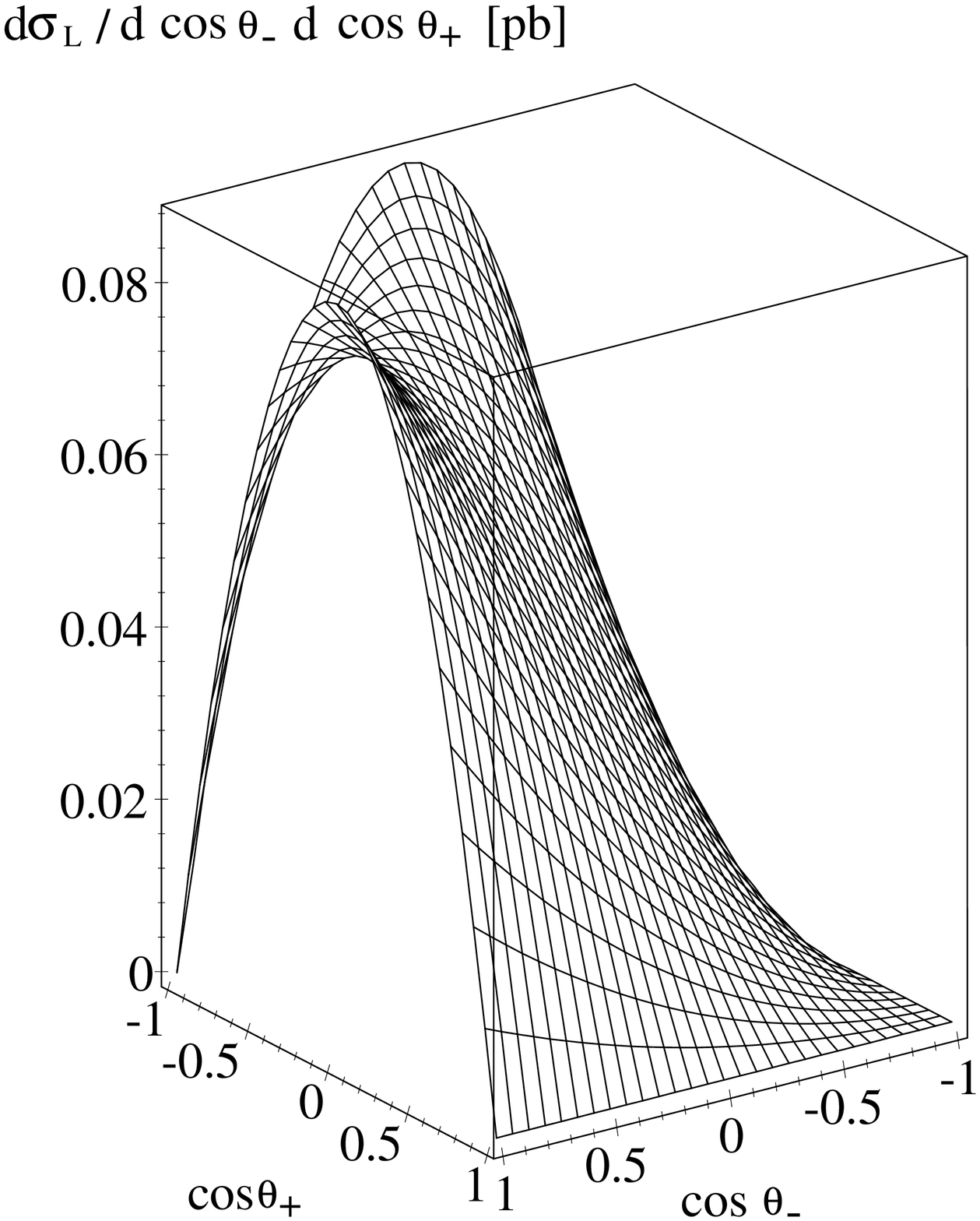}}
\vspace*{-0.3cm}
\centerline{Figure 1}
\end{figure}
\end{minipage}
\begin{minipage}[c]{3cm}
\quad
\end{minipage}
\begin{minipage}[r]{5cm}
\begin{figure}[H]
\centerline{\epsfysize=10cm\epsffile{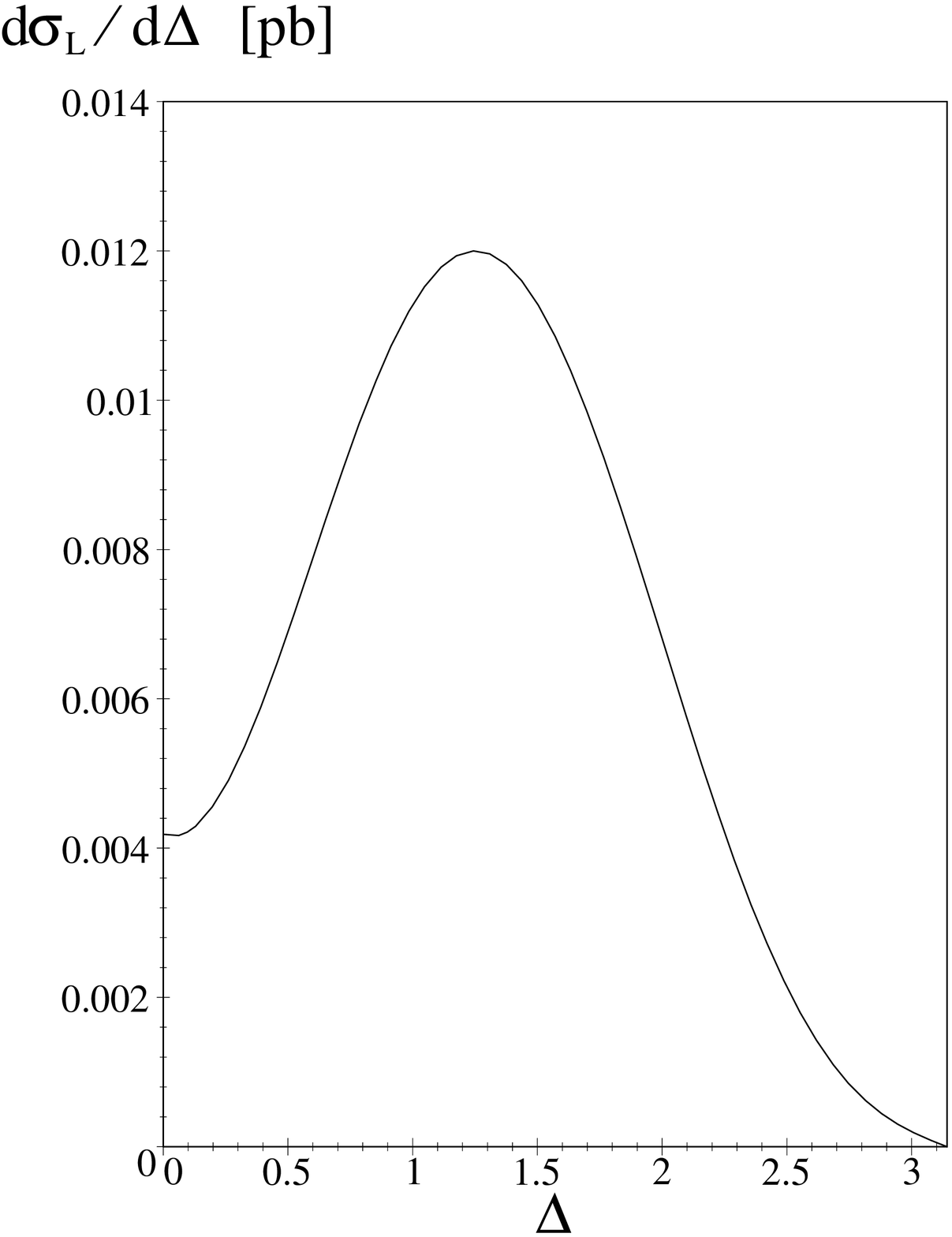}}
\vspace*{-0.3cm}
\centerline{Figure 2}
\end{figure}
\end{minipage}

\vspace{0.6cm}
\hspace*{-2.7cm}
\begin{minipage}[l]{9cm}
\begin{figure}[H]
\centerline{\epsfysize=10cm\epsffile{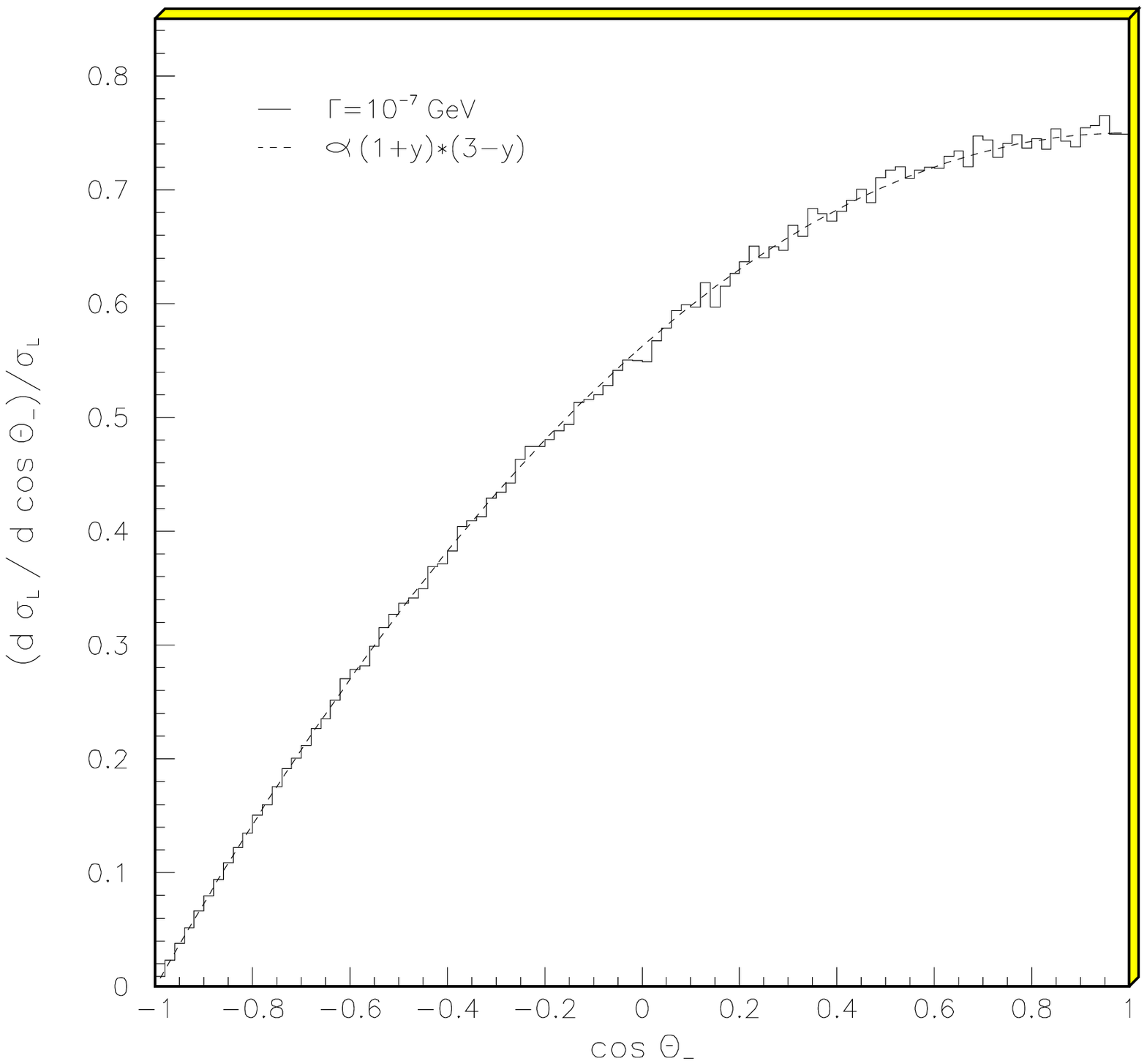}}
\end{figure}
\end{minipage}
\begin{minipage}[r]{9cm}
\begin{figure}[H]
\centerline{\epsfysize=10cm\epsffile{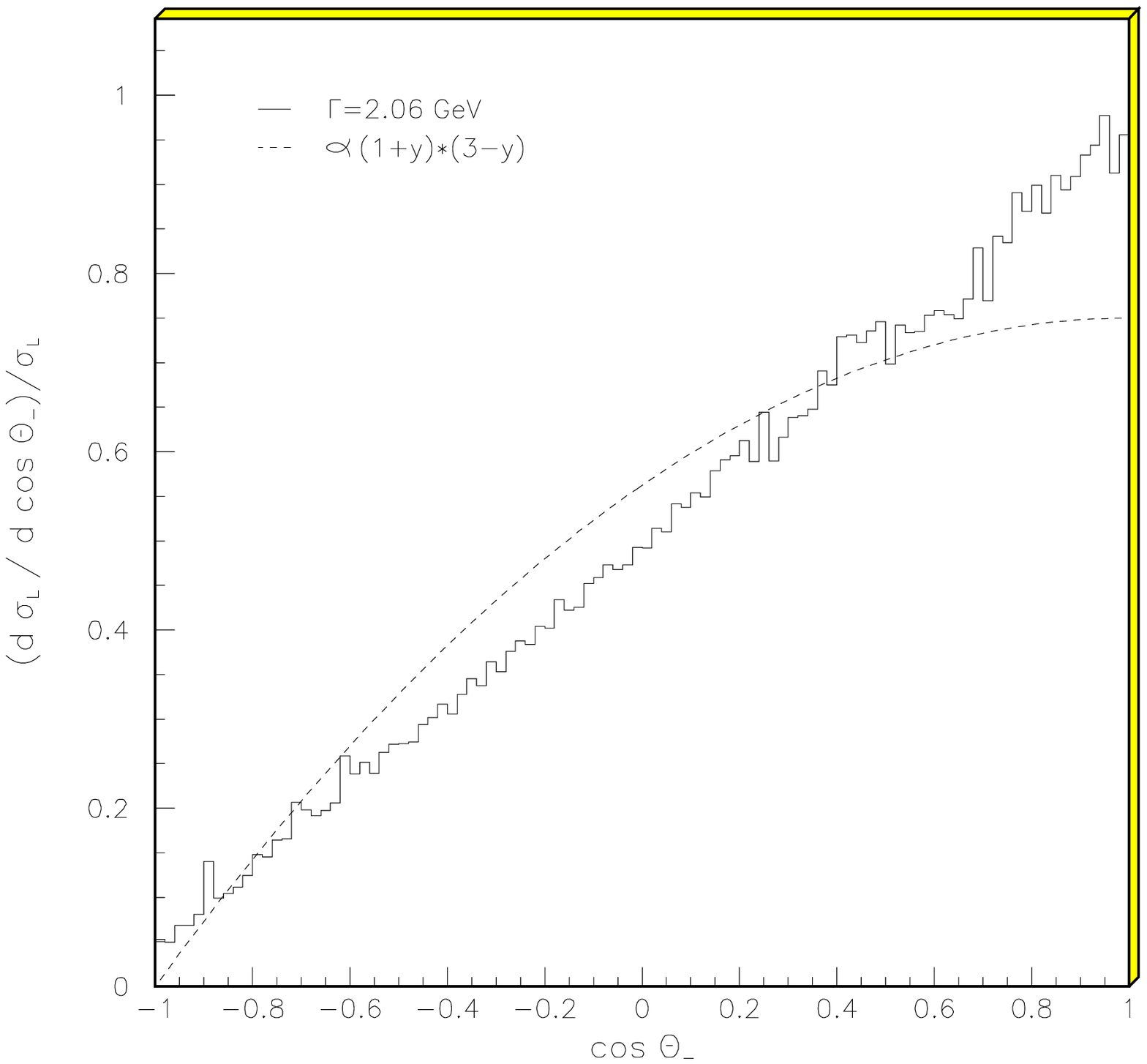}}
\end{figure}
\end{minipage}
\hspace*{-1cm}
\centerline{Figure 3}

\newpage

\vspace*{-3.5cm}
\begin{minipage}[l]{5cm}
\begin{figure}[H]
\centerline{\epsfysize=10cm\epsffile{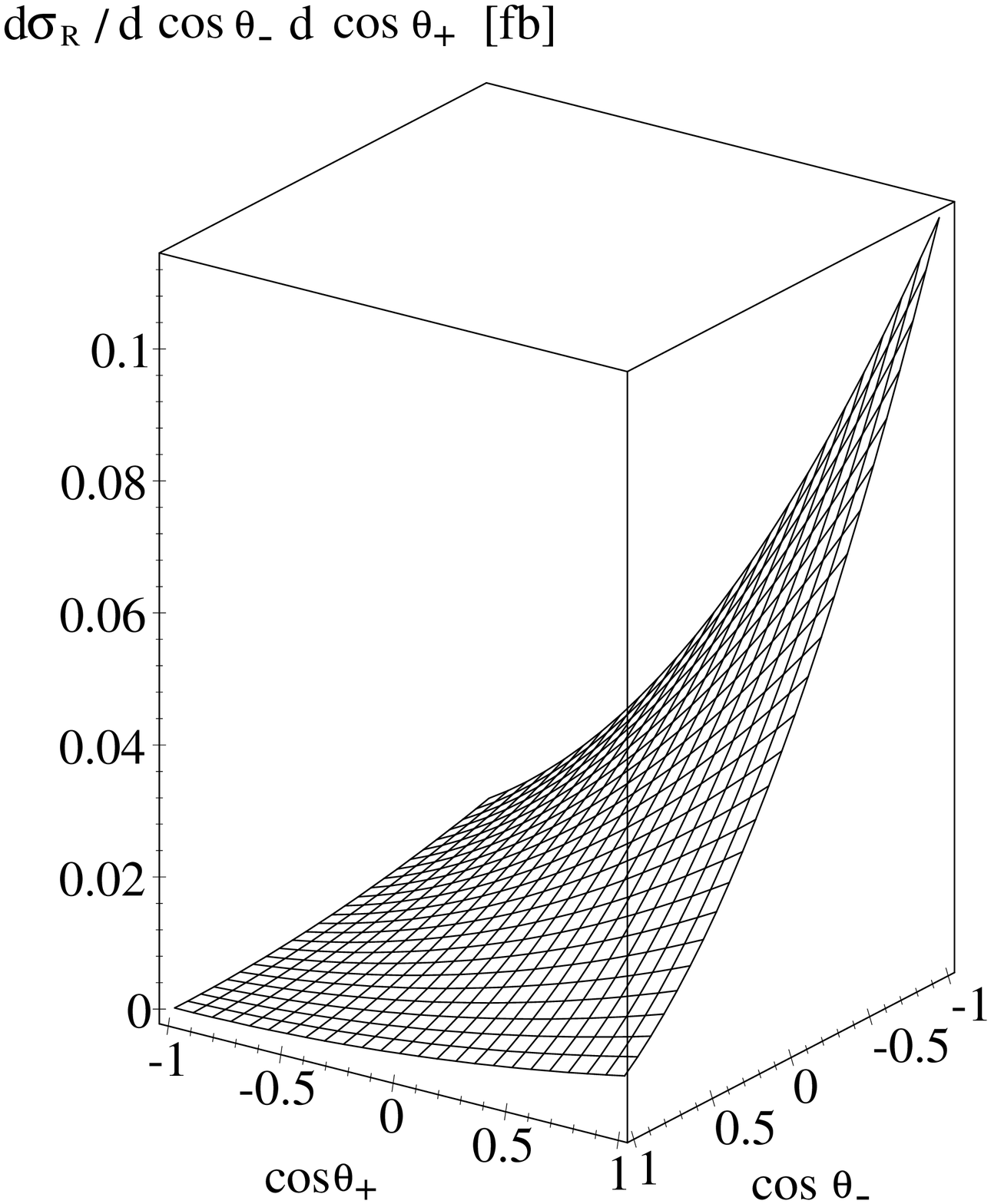}}
\vspace*{-0.3cm}
\centerline{Figure 4}
\end{figure}
\end{minipage}
\begin{minipage}[c]{3cm}
\quad
\end{minipage}
\begin{minipage}[r]{5cm}
\begin{figure}[H]
\centerline{\epsfysize=10cm\epsffile{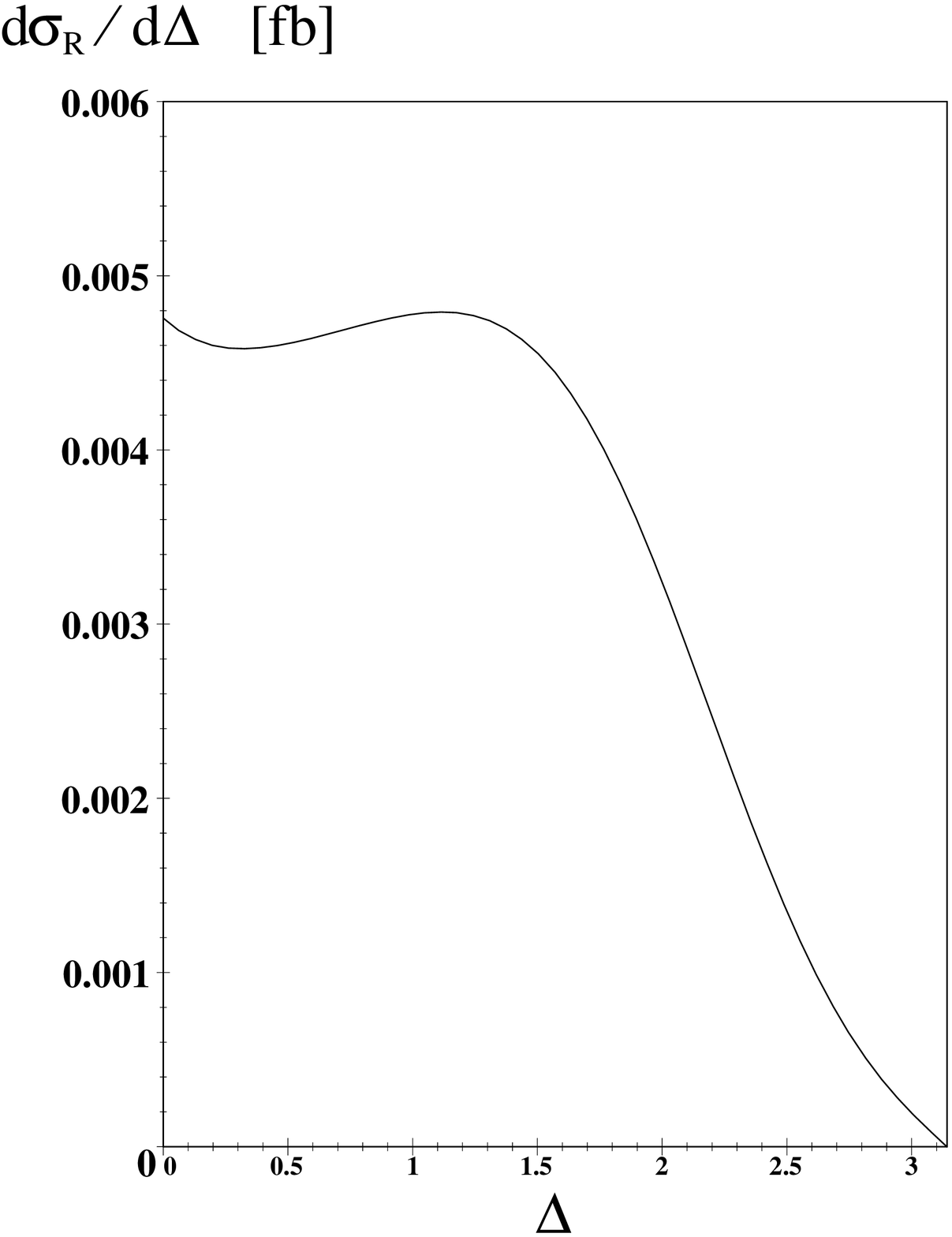}}
\vspace*{-0.3cm}
\centerline{Figure 5}
\end{figure}
\end{minipage}

\vspace{0.6cm}
\hspace*{-2.7cm}
\begin{minipage}[l]{9cm}
\begin{figure}[H]
\centerline{\epsfysize=10cm\epsffile{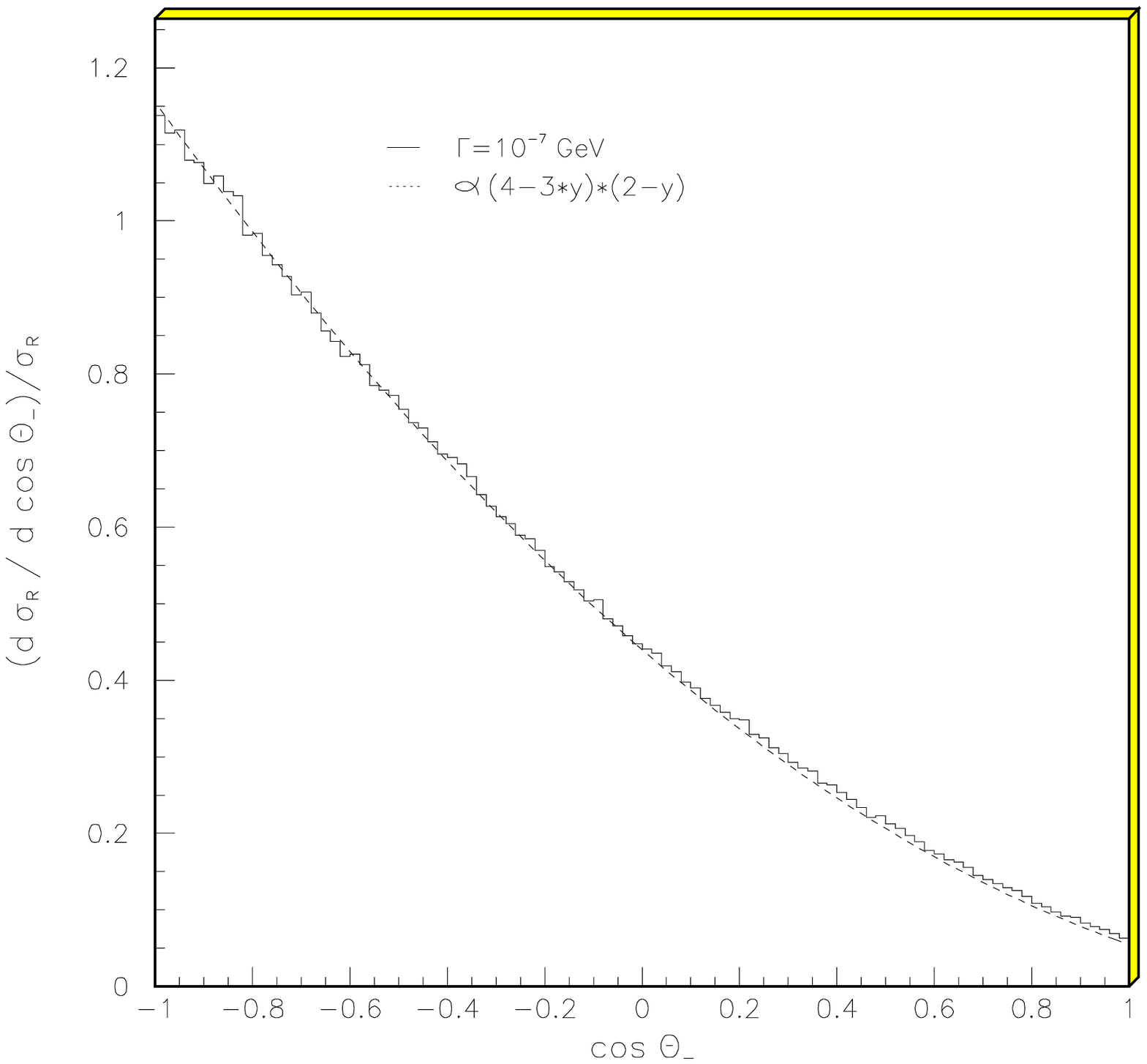}}
\end{figure}
\end{minipage}
\begin{minipage}[r]{9cm}
\begin{figure}[H]
\centerline{\epsfysize=10cm\epsffile{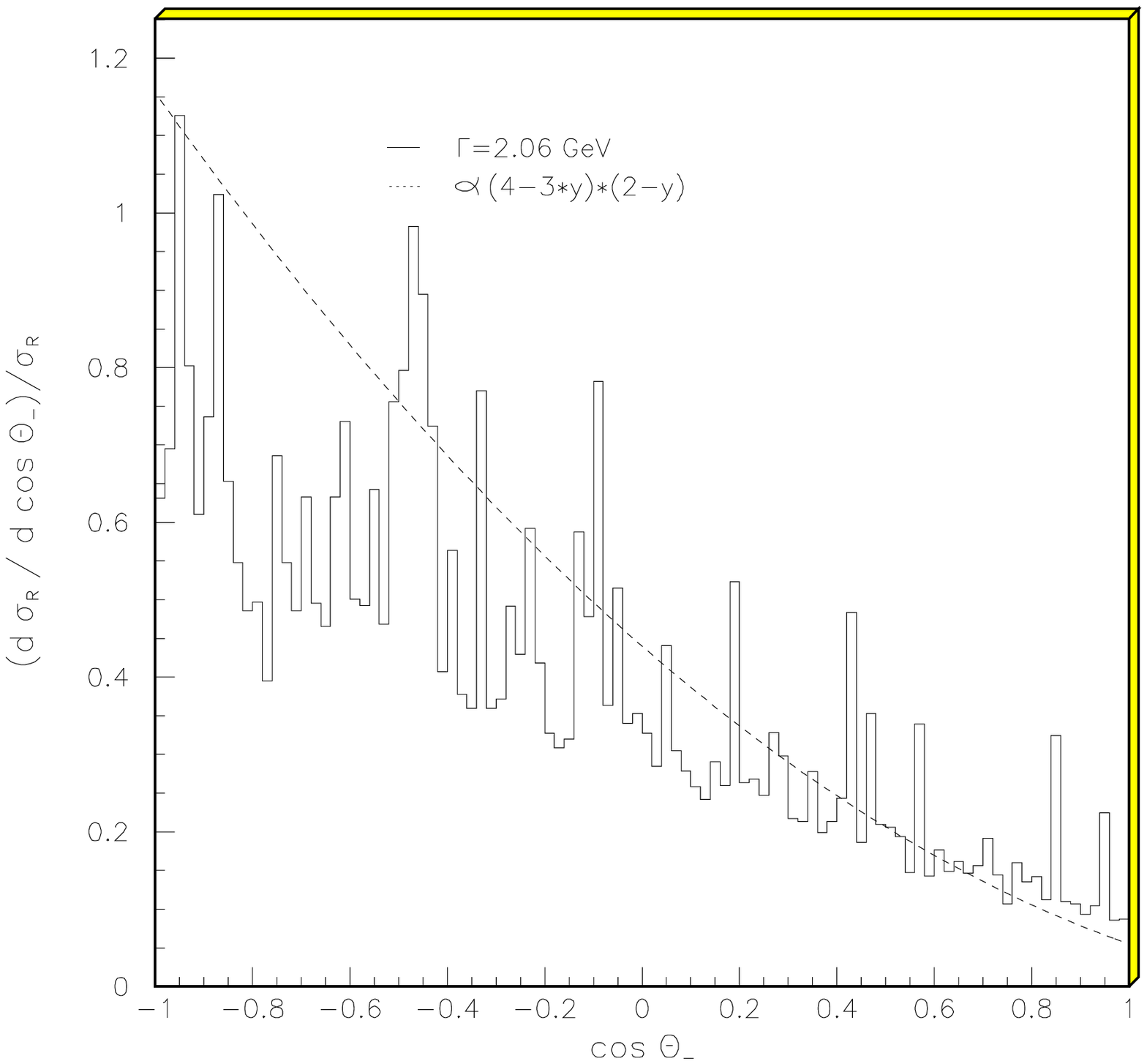}}
\end{figure}
\end{minipage}
\hspace*{-1cm}
\centerline{Figure 6}

\newpage

\end{document}